\documentclass[notitlepage,aps,prd,amsmath,floats,floatfix,
twocolumn,
superscriptaddress,nofootinbib,showpacs]{revtex4-1}

\usepackage{amsfonts,amsmath,units,wasysym,epsfig,graphicx,verbatim,color,subfigure,graphicx}
\usepackage{amsmath}
\usepackage{amssymb}
\usepackage{amsfonts}
\usepackage{bm}
\usepackage{color}
\usepackage[normalem]{ulem}
\graphicspath{{Images/}}

\usepackage{lipsum}
\usepackage{natbib}
\usepackage{color,soul}
\usepackage{hyperref}
\usepackage{marginnote}
\usepackage{braket}
\usepackage[none]{hyphenat}
\usepackage{bm}
\usepackage{comment}
\usepackage{multirow}
\usepackage{array}

\definecolor{green1}{RGB}{17,119,51}
\definecolor{yellow1}{RGB}{221,204,119}
\definecolor{pink1}{RGB}{204,102,119}
\definecolor{blue1}{RGB}{68,119,170}
\definecolor{blue2}{RGB}{136,204,238}

\newcolumntype{P}[1]{>{\centering\arraybackslash}p{#1}}

\begin{document}

\newcommand{\SU}{\affiliation{Department of Physics, Syracuse University, Syracuse, New York 13244, USA}}
\newcommand{\princeton}{\affiliation{Department of Astrophysical Sciences, Princeton University, Princeton, NJ 08544}}
\newcommand{\ias}{\affiliation{Institute for Advanced Study, 1 Einstein Dr, Princeton NJ 08540}}


\title{Detection Prospects of Core-Collapse Supernovae with Supernova-Optimized Third-Generation Gravitational-wave Detectors}

\author{Varun Srivastava}\SU
\author{Stefan Ballmer}\SU
\author{Duncan~A.~Brown}\SU
\author{Chaitanya~Afle}\SU
\author{Adam Burrows}\princeton
\author{David Radice}\princeton\ias
\author{David Vartanyan}\princeton

\date{\today}


\begin{abstract}
We optimize the third-generation gravitational-wave detector to maximize the range to detect core-collapse supernovae. Based on three-dimensional simulations for core-collapse and the corresponding gravitational-wave waveform emitted, the corresponding detection range for these waveforms is limited to within our galaxy even in the era of third-generation detectors. The corresponding event rate is two per century. We find from the waveforms that to detect core-collapse supernovae with an event rate of one per year, the gravitational-wave detectors need a strain sensitivity of 3$\times10^{-27}~$Hz$^{-1/2}$ in a frequency range from 100~Hz to 1500~Hz. We also explore detector configurations technologically beyond the scope of third-generation detectors. We find with these improvements, the event rate for gravitational-wave observations from CCSN is still low, but is improved to one in twenty years.
\end{abstract}

\maketitle

\section{Introduction}\label{sec:intro}
The Advanced LIGO \cite{aasi2015advanced} and VIRGO \cite{acernese2014advanced}  gravitational-wave detectors observed signals from the coalescence of over ten binary black holes (BBH) and one binary neutron star merger (BNS) \cite{abbott2016observation, abbott2016gw151226, abbott2017gw170817, GWTC1, venumadhav2019new, nitz20191} by the end of their second science run.
Core-collapse supernovae (CCSN) are a potential astrophysical source of gravitational waves that could be detected by interferometric detectors. The gravitational waves are generated deep in the star, at the collapsing core, and are emitted untouched by the outer envelopes. They contain vital information about the interior of the star and about the core-collapse process, which is not present in the electromagnetic counterpart of the emitted radiation. We can infer various physical parameters such as the nuclear equation of state, rotation rate, pulsation frequencies, etc. from the gravitational wave signal of a CCSN once it has been detected \cite{torres2019universal,richers2017equation,abdikamalov2014measuring}. However, gravitational waves from CCSN are yet to be observed \cite{abbott2016first,abbott2019all}. The inferred sensitivity of the aLIGO-VIRGO network to detect CCSN ranges from a few kiloparsecs (kpc) to a few megaparsecs (Mpc) \cite{gossan2015}. The range of a few megaparsecs in \citet{gossan2015} corresponds to extreme emission models which assume properties of stars which are unlikely to occur in astrophysical scenarios. The smaller sensitive range of a few kiloparsecs to CCSN along with low CCSN rates within galaxies leads to a low gravitational-wave detection probability from CCSN \cite{li2011nearby2, li2011nearby3, Graham_2012, xiao2015_11Mpc, 2019ApJ...876L...9R}.

The gravitational radiation from CCSN depends on a complex interplay of general relativity, magneto-hydrodynamics, nuclear, and particle physics. The burst signal, therefore, does not have a simple model, and we have to use numerical simulations to understand its structure. Numerical simulations also help in understanding the frequency content of the gravitational wave signal which is crucial in determining the parameters to tune future detectors towards supernovae. 

The three-dimensional (3D) simulations of core-collapse supernovae reveal that their gravitational-wave signatures are broadband with frequencies ranging from a few hertz to a few thousand hertz.  The time-changing quadrupole moment of the emitted neutrinos occupies the few Hertz to ten Hertz range, while the higher frequencies are associated with the prompt convection and rotational bounce phase, the proto-neutron-star (PNS) ringing phase, and turbulent motions. \citet{murphy:09} and \citet{2018ApJ...861...10M} demonstrated that the excitation of the fundamental g- and f-modes of the PNS can be a dominant component and that much of the gravitational wave energy emitted is associated with such PNS oscillations.  The frequency ramp with time after the bounce of the latter is a characteristic signature of CCSN and will reveal the inner dynamics of the residual PNS core and supernova phenomenon once detected.  There now exist in the literature numerous 3D CCSN models that map out the gravitational-wave signatures expected from CCSN \citep{kuroda:14,andresen:17,yakunin:17,Kuroda:2017trn,2019ApJ...876L...9R,powell2018gravitational}.  For this study, we focus on the extensive suite of 3D waveforms found in \citet{2019ApJ...876L...9R}.

In our work, we optimize the design prospects of a third-generation Cosmic-Explorer-like detector to detect gravitational wave signals from CCSN and discuss the astrophysical consequences. We focus on the prospects for detection of non-rotating or slowly rotating stars since they are likely to be astrophysically more likely \cite{scalo1986stellar}. We first review the detection ranges for the second-generation detectors. A significant amount of power is emitted by CCSN within the gravitational-wave frequency range 500~Hz to 1500~Hz. Therefore, in order to improve the sensitivity of gravitational wave detectors to CCSN, we need to tune the detector parameters to increase the sensitivity in this bandwidth. With the present models of likely gravitational wave emission from CCSN \citep{2019ApJ...876L...9R}, we find that the detectable range with a supernovae-optimized Cosmic-Explorer-like third generation detector is still only up to a hundred kiloparsecs. The detector range is therefore limited to CCSN that occur within our galaxy. The corresponding event rate is approximately two per century \cite{adams2013observing,cappellaro1997rate,timmes1997gamma,pierce1994hubble,tammann1994galactic}. However, the supernovae-optimized detector would improve the signal-to-noise ratio (SNR) for the galactic sources by approximately 25$\%$ as compared to the Cosmic-Explorer. For completeness, we also discuss the strain requirements in a detector to achieve CCSN event rates of the order of one per year. To this end, we address the fundamental sources of noise that limit our sensitivity to achieve this desired strain. 



\begin{table*}
  	\begin{center}
  	\begin{tabular}{|c|c|c|}
  		\hline
        \ Distance & Type-II CCSN rate (per century) & References \\
        \hline
        \ Milky way (D $<$ 30 kpc) & 0.6-2.5 & \cite{adams2013observing,cappellaro1997rate,timmes1997gamma,pierce1994hubble,tammann1994galactic} \\
        \hline 
        \ M31 or Andromeda (D = 770 kpc) & 0.2-0.83 & \cite{capaccioli1989properties,1994ApJS...92..487T,freedman1990empirical,timmes1997gamma,tammann1994galactic} \\
        \hline
        \ M33 (D = 840 kpc) & 0.62 & \cite{timmes1997gamma,tammann1994galactic} \\
        \hline
        \ Local Group ( D $<$ 3 Mpc) & 9 & \cite{mattila2001supernovae,tammann1994galactic} \\
        \hline 
        \ Edge of Virgo Super-cluster (D $<$ 10 Mpc) & 47 & \cite{Botticella2012,mattila2012core,kistler2011core} \\
        \hline 
        \ Virgo-cluster (D $<$ 20 Mpc) & 210 & \cite{mattila2012core,van1991galactic} \\
        \hline 
  	\end{tabular}
  	\end{center}
    \caption{The cumulative rate of CCSN in out local universe. To achieve a detection rate of one per year, assuming a 100$\%$ duty cycle of the gravitational wave detector, we need a strain sensitivity to have a CCSN reach of the order of 10 Mpc.}
    \label{tab:SNrate}
\end{table*}


\section{Gravitational waves from CCSN}\label{sec:GWsSN}


In this paper, we use the 3D simulations for CCSN from \citet{2019ApJ...876L...9R} for progenitors with Zero Age Main Sequence (ZAMS) masses $9, 11, 19, 25$ and $60 M_{\odot}$. The simulations were performed with state-of-the-art neutrino-radiation hydrodynamics using the Eulerian radiation-hydrodynamics code \texttt{FORNAX} \citep{2019ApJS..241....7S, skinner2016should}. \citet{2019ApJ...876L...9R} used progenitors from \citet{sukhbold2016core,sukhbold2018high} with solar metalicity. The comparison 2D models were taken from \citet{2018ApJ...861...10M}.

\begin{figure*}
    \vspace{-2.2cm}
	\centering
	\includegraphics[width=0.95\textwidth]{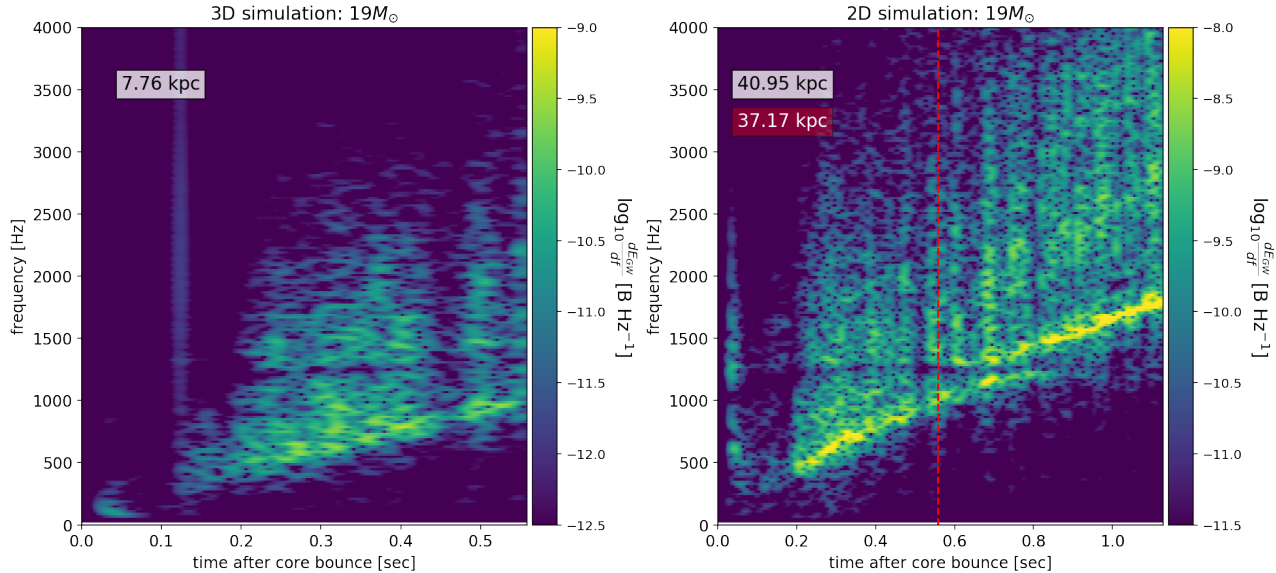}
	\vspace{-2.7cm}
	\caption{Spectrograms of gravitational-wave waveforms from 3D (left column) and 2D (right column) simulations of $19 M_{\odot}$ progenitor. The number on the top left corner each plot with white background is the distance for which these GW signals have an optimal SNR of 8. For the 2D simulations, we recalculate this distance (shown on red background)  by truncating the waveform at the end-time of the corresponding 3D simulation. The red vertical dashed line shows the truncation time.}
	\label{fig:9_spectograms}
\end{figure*}

Fig.~\ref{fig:9_spectograms} shows the spectrograms of the waveforms obtained from the simulation for the $19 M_{\odot}$ progenitor. The left column shows the spectrogram of the waveform from the 3D simulation, while the right column shows the spectrogram of the waveform from the 2D simulation. The red vertical dashed line in the right column represents the simulation time of the 3D waveform. For simulations of the same ZAMS mass, both the 2D waveforms and the 3D waveforms show similar behavior in the time-frequency plane. We can see the prompt convection signal for the first $\sim 10$ milliseconds after the core bounce, followed by the characteristic g/f-mode ring up of the proto-neutron star (PNS) \cite{Morozova_2018_2D}. For the 2D waveforms, the frequency ranges from $\sim20$ Hz to $\sim2000$ Hz. The g/f-mode ring up of the PNS starts around 200 milliseconds after the core bounce at a frequency of $\sim 500$ Hz, and $1$ sec after the core bounce reaches $\sim 1500$ Hz. For the waveforms obtained from the 3D simulations, the frequency ranges up to $\sim 1000$ Hz. This is because the 3D simulations end earlier ($0.4-1.0$~sec after core bounce). 

\begin{table*}
  	\begin{center}
  	
  	\begin{tabular}{|P{3cm}|P{1.5cm}|P{1.5cm}|P{2cm}|P{2cm}|P{2cm}|P{2cm}|P{2cm}|}
  		\hline
        \ \multirow{2}{*}{ZAMS Mass} & \multicolumn{3}{|p{5cm}|}{\centering Optimal distance (kpc)} &  \multicolumn{2}{|p{4cm}|}{\centering Normalized $\sigma^2$ (10Hz-450Hz)} &  \multicolumn{2}{|p{4cm}|}{\centering Normalized $\sigma^2$ (450Hz-2000Hz)} \\
            \cline{2-8}
            & 3D & 2D & 2D truncated & 3D & 2D & 3D & 2D \\
        \hline
        \ $9 M_{\odot}$ & $2.43$ & $15.51$ & $15.46$ & $0.342$ & $0.232$ & $0.658$ & $0.767$ \\
        \hline 
        \ $11 M_{\odot}$ & $5.87$ & $31.96$ & $26.68$ & $0.154$ & $0.058$ & $0.845$ & $0.941$ \\
        \hline 
        \ $11 M_{\odot}$ (w/o MB) & $5.99$ & $28.78$ & $26.04$ & $0.131$ & $0.099$ & $0.869$ & $0.9$ \\
        \hline 
        \ $19 M_{\odot}$ (w/o MB) & $7.75$ & $40.61$ & $37.18$ & $0.120$ & $0.074$ & $0.880$ & $0.925$ \\
        \hline 
        \ $25 M_{\odot}$ & $13.35$ & $48.26$ & $40.09$ & $0.12$ & $0.069$ & $0.88$ & $0.931$ \\
         \hline 
        \ $60 M_{\odot}$ & $9.63$ & $48.79$ & $36.30$ & $0.211$ & $0.065$ & $0.790$ & $0.935$ \\
        \hline 
  	\end{tabular}
  	\end{center}
    \caption{The optimal distances for aLIGO and $\sigma^2$ for a flat PSD in the frequency bandwidths 10Hz - 450Hz and 450Hz - 2000Hz for waveforms.}
    \label{tab:opt_dist_sigma}
\end{table*}

We calculate the optimal distance for each of these waveforms, as defined below \cite{finn1993observing}:
\begin{equation}
    \label{eq:optD}
    d_{opt} = \frac{\sigma}{\rho^*} = \frac{1}{\rho^*}\left[ 2  \int_{f_{\text{low}}}^{f_{\text{high}}} df \frac{\Tilde{h}(f)\Tilde{h}^*(f) }{S_{h} (f)}\right]^{\frac{1}{2}}
\end{equation}
where $S_{h} (f)$ is the power spectral density (PSD) of the detector, $\rho^* = 8$ is the optimal signal-to-noise ratio (SNR) and the limits over the integral are defined by $f_{\text{low}}$ and $f_{\text{high}}$. We set the lower frequency cutoff, $f_{\text{low}} = 10$ Hz and use \texttt{aLIGOZeroDetHighPower} \cite{lalsuite} as PSD for aLIGO to compute the optimal distances for all the waveforms, which are shown in Table \ref{tab:opt_dist_sigma}. For aLIGO, the average distances for waveforms from 3D simulations are $\sim 8$ kpc, while the average distances for corresponding 2D numerical simulations are $\sim 35.5$ kpc. The 3D simulations have shorter times with respect to the 2D simulations, so we truncate the 2D simulations at the same corresponding times to compare the optimal distances. In doing so, the average optimal distance for the waveforms from the 2D simulations is $\sim 30$ kpc. We find that the 2D waveforms are, on an average, $\sim 4$ times louder than the 3D waveforms. Therefore, we will only use the waveforms from 3D simulations to tune the third generation detectors for CCSN and calculate ranges.

Table \ref{tab:opt_dist_sigma} also shows the optimal signal-to-noise (SNR) $\sigma^2$ of the waveforms in two frequency bandwidths : 10Hz - 450Hz and 450Hz - 2000Hz. These $\sigma^2$ values have been calculated using a flat PSD (see section \S\ref{sec:FutureDet}), so that we can infer the distribution of the frequency content of the waveforms without being biased by the noise curves of any detector. We can verify from the spectrograms that almost all of the frequency content is below $2000$ Hz. We find that the ratio of $\sigma^2$ in the range 10Hz - 450Hz to that in range 450Hz - 2000Hz is $\sim 0.2$ for 3D simulations while for 2D simulations it is $\sim 0.1$. This implies that $\sim 80 \%$ of the content of the waveforms is in the frequency range  450Hz - 2000Hz. This is crucial since in Secs \S\ref{sec:Pheno} and \S\ref{sec:SN3GDet}, we tune the detector parameters to increase the sensitivity in this frequency range.

In section \S\ref{sec:Pheno}, we define a phenomenological CCSN waveform which is derived from the 3D numerical waveforms. We maximize the range of the phenomenological supernovae waveform (see Fig. \ref{fig:ModelCCSN}) with a third-generation Cosmic-Explorer-like detector. We use \textsc{GWINC} to estimate the noise floor for different detector parameters \cite{gwinc}. The maximized range achieved can then be translated into the corresponding event rate of CCSN, as summarized in table \ref{tab:SNrate} (assuming a 100$\%$ detector duty-cycle). 

We use the waveforms from \citet{2019ApJ...876L...9R} to compare the ranges of different waveforms of CCSN using the Einstein Telescope (ET), the Cosmic Explorer (CE) and the Supernovae-Optimized detector (SN-Opt). In section \ref{sec:FutureDet}, we invert the problem to calculate the strain requirements of a {\it hypothetical} detector to achieve an event rate of the order of one in two years or in the terms of distances -- has a range of the order of 10 Mpc for gravitational-wave signals from CCSN. Lastly, we consider in section \S\ref{sec:FutureDet} detector configurations beyond the third-generation detectors (Hypothetical) and find the ranges for different numerical waveforms of CCSN.

\section{Defining a Representative Supernovae Gravitational-Wave Waveform}\label{sec:Pheno}
To maximize the detectable range for CCSN in a given detector configuration, we need a reference CCSN waveform that captures the broad features of supernovae waveform. The reference waveform must have the strain amplitude and spectral features similar to any supernovae waveform. We use the waveforms from the 3D simulations of core-collapse \cite{burrows2019three,2019ApJ...876L...9R} to generate a phenomenological model that captures the broad range of features of core-collapse supernovae waveform. We generate the phenomenological waveform to average out the power emission features from different numerical waveforms so that features in any one of the waveforms do not affect the results of the study. Thereby, the phenomenological waveform provides a model-independent approach.

We construct the phenomenological waveform by a sum of sine-Gaussian bursts. A sine-Gaussian can be defined with three parameters, the central frequency $f_{o}$, the quality factor or the sharpness of the peak $Q$ and the amplitude scale $h_{o}$. The frequency domain representation of a sine-Gaussian can be expressed with these parameters as

\begin{equation}\label{eq:pheno}
    \Tilde{s}(f) = \frac{h_{o}}{4\sqrt{\pi}} \frac{Q}{f_{o}} e^{-\frac{(f-f_{o})^2 Q^2}{4f^2}}
\end{equation}

The different frequencies are used to model different spectral features of the core-collapse waveform. We choose central frequencies $f_{o}^{i}$ for sine-Gaussian using the numerical waveforms from 3D simulations of core-collapse. We choose, by hand, five distinct central frequencies $f_{o}^{i}$ which correspond to peak emission in the numerical waveforms. We limit ourselves to five distinct values of frequencies in order to avoid over-fitting the sine-Gaussian phenomenological waveform to the numerical waveforms. 
We note that the supernovae waveforms have emission at higher frequencies but they are much lower in amplitude. Therefore, for the purposes of optimization, we limit ourselves to an upper limit of 2kHz in the phenomenological waveform. 

To build the phenomenological waveform, we divide the frequency domain into four bins ranging from -- 10~Hz to 250~Hz, 250~Hz to 500~Hz, 500~Hz to 1~kHz and 1~kHz to 2~kHz. For each of the chosen central frequencies $f_{o}^{i}$, the quality factor $Q^i$ and the amplitude $h_{o}^{i}$ are chosen so as to minimize the error in the normalized power in the four different bins of frequencies above. The error in the normalized power in each bin is then added in quadrature for different waveforms and is given by

\begin{equation}
    \Delta e = \sqrt{ \frac{1}{N-1} \sum_{i}^{N} (\mathrm{Model}_{f_{low}}^{f_{high}} - \mathrm{NR}_{f_{low}}^{f_{high}})^2 }
\end{equation}

\begin{figure}
    \centering
    \includegraphics[width=0.45\textwidth]{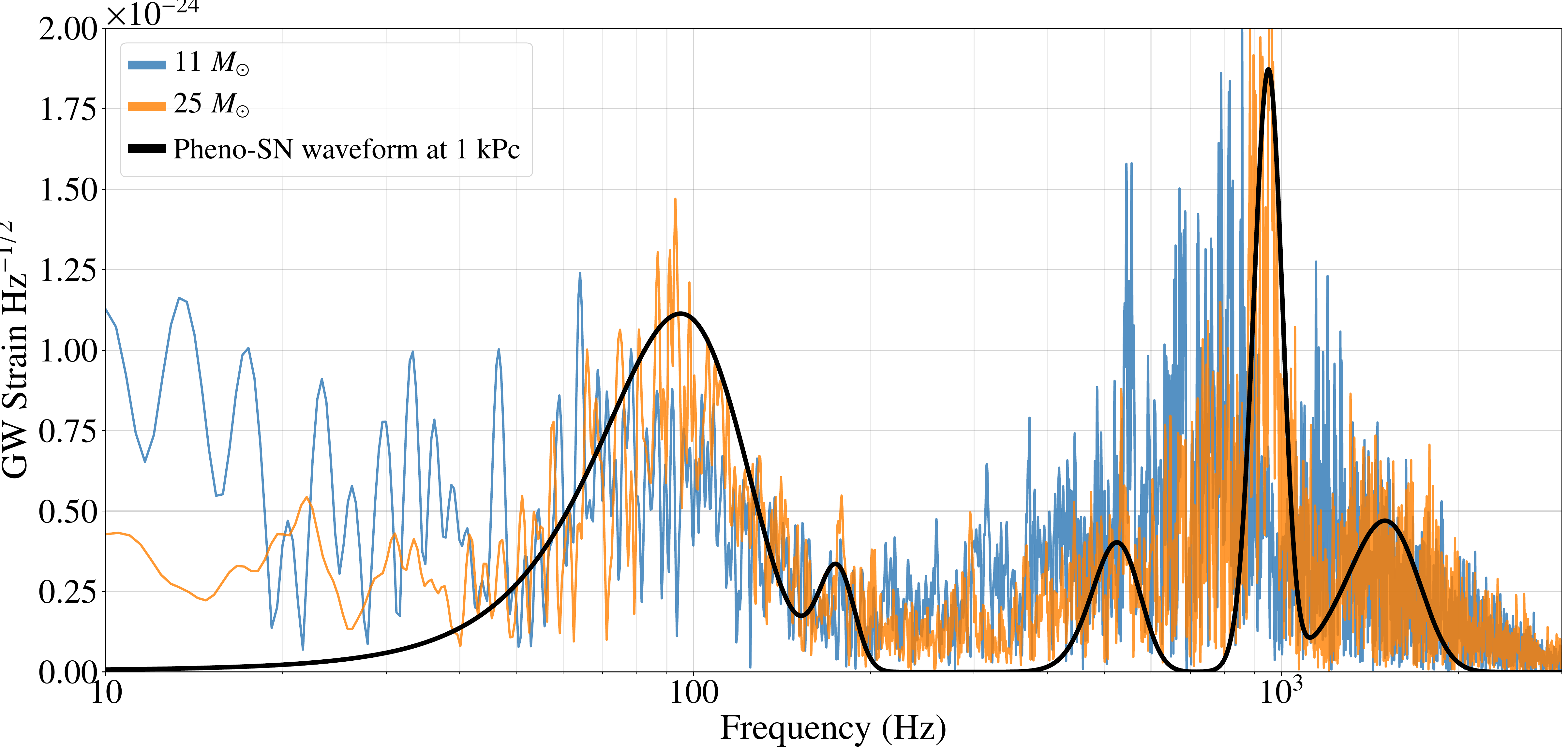}
    \caption{The figure shows the phenomenological waveform used as a representative for gravitational wave emission from CCSN. The waveform is constructed by using five sine-Gaussian bursts with different central frequencies $f_o = $ 95, 175, 525, 950 and 1500~Hz. The quality factor and the amplitude at each central frequency are then derived by minimizing the normalized power emitted in four different bins of frequency from 10~Hz to 250~Hz, 250~Hz to 500~Hz, 500~Hz to 1000~Hz and 1000~Hz to 2000~Hz. The overall amplitude of the phenomenological waveform is not calculated by the fit and can be rescaled. We are interested in the broad features in frequency in different waveforms which is effectively captured in the phenomenological waveform. }
    \label{fig:ModelCCSN}
\end{figure}

This approach gives us a simple but robust gravitational waveform, free from the parameter degeneracies but capturing the features of gravitational wave radiation from CCSN. We will use this to perform optimization and maximize the range for this waveform and thus for CCSN. The errors in the different frequency bins ranging from 10Hz to 250Hz, 250Hz to 500Hz, 500Hz to 1kHz and 1kHz to 2kHz is 3$\%$, 9$\%$, 2$\%$ and 19$\%$ respectively. The higher error in the last frequency bin is by the construction of the phenomenological waveform and is added to incorporate the features persistent in the 2D waveforms which show higher emissions in this frequency range discussed in section \S\ref{sec:GWsSN}. Fig. \ref{fig:ModelCCSN} shows the phenomenological waveform constructed. We incorporate this waveform as a reference supernovae signal within \texttt{GWINC} \cite{gwinc}. The ranges, horizon, and reach for the phenomenological waveform can then be calculated by solving for distance $D$ which would rescale the waveform in equation \ref{eq:pheno} as $1/\mathrm{D}$.

In each of the subsequent sections, we go back to each of the numerical waveforms and recompute the ranges achieved with all the different detector designs considered in our study. 

\section{Optimizing SN detectability for 3G detectors}\label{sec:SN3GDet}
We use the phenomenological gravitational-wave waveform for CCSN to explore detector configurations that optimize the Cosmic Explorer detector's sensitivity to CCSN. To avoid overemphasis on any particular frequency chosen in the phenomenological waveform, we down weight narrow-band configurations during the process of optimization. We also avoid narrow-band designs so that the optimized detector's sensitivity to BNS is greater then 1~Gpc. We will explore the narrow-band configurations with a different approach discussed in section \S\ref{subsec:NBDetDes}

\subsection{Broadband configuration tuned for Supernovae} \label{subsec:BB_CCSNDet}
Quantum noise is the predominant source of noise which limits the performance of the gravitational-wave detector. Radiation pressure noise limits the detector sensitivity at low frequencies and shot noise limits sensitivity at high frequencies \cite{aasi2015advanced, acernese2014advanced, buonanno2003quantum}. In our study, we use the design parameters of Cosmic Explorer \cite{CEdesignpaper} as the starting point. For the purposes of optimization, we choose the Cosmic Explorer rather than the Einstein Telescope as the former has a better noise performance at frequencies which are relevant to CCSN. We optimize over the length of the signal recycling cavity ($\mathrm{L}_{src}$) and the transmissivity of the signal recycling mirror ($\mathrm{T}_{srm}$) to maximize the CCSN detection range. The quantum resonant sidebands can be tuned with these parameters and we exploit this behavior for supernovae tuning similar to the approach used by Buonanno~et~al. \cite{BunnChen2004} and Martynov~et~al. \cite{martynov2019exploring}.

We also study, the effect of the length of the arm cavity ($\mathrm{L}_{arm}$) on supernovae sensitivity. We use Markov Chain Monte Carlo sampling \cite{hastings1970monte} and particle swarm optimization \cite{PSO1995} to search the parameter space and maximize the range for the phenomenological waveform for a broadband detector. During the process of maximizing the range, we down-weight the narrow-band configurations with two constraints for sample points. First, the reflectivity of the signal recycling cavity $\mathrm{T}_{srm} > $ 0.01. Second, the given detector configuration must have a optimal distance for binary neutron stars systems ($m_1 = m_2 = 1.4~M_{\odot}$ and $s_{1z} = s_{2z} = 0$) to be greater than 1 Gpc. By doing so, we ensure that the detector's sensitivity is not lost for compact binaries.

The strain sensitivity improves as the square root of the arm length of the detector as long as the gravitational-wave frequency ($\Omega$) is much less than the free spectral range ($\mathrm{f}_{FSR}$) of the Fabry-Perot cavity. The strain sensitivity of the detector does not always improve by scaling the detector as other fundamental sources of noise also change by scaling the length of the detector \cite{dwyer2015gravitational}. As the gravitational wave spectrum of supernovae has some power in a few kilohertz range, we allow the arm length to vary independently similar to the analysis by \cite{Miao2018,martynov2019exploring}. Our simulations indicated the optimal length to be close to 40 km, the upper bound value allowed for the length parameter. As a result, we set the length of the arm cavity to 40 km. For a 40 km arm length, the $\mathrm{f}_{FSR}$ is 3750 Hz. The sensitivity of the detector is limited by the $\mathrm{f}_{FSR}$, any further increase in the length of the arms will reduce the $\mathrm{f}_{FSR}$, resulting in the loss in sensitivity to CCSN, where the gravitational wave spectrum persists up to a few kilohertz.

\begin{figure}
    \hspace*{-0.7cm} 
	\centering
	\includegraphics[width=0.49\textwidth]{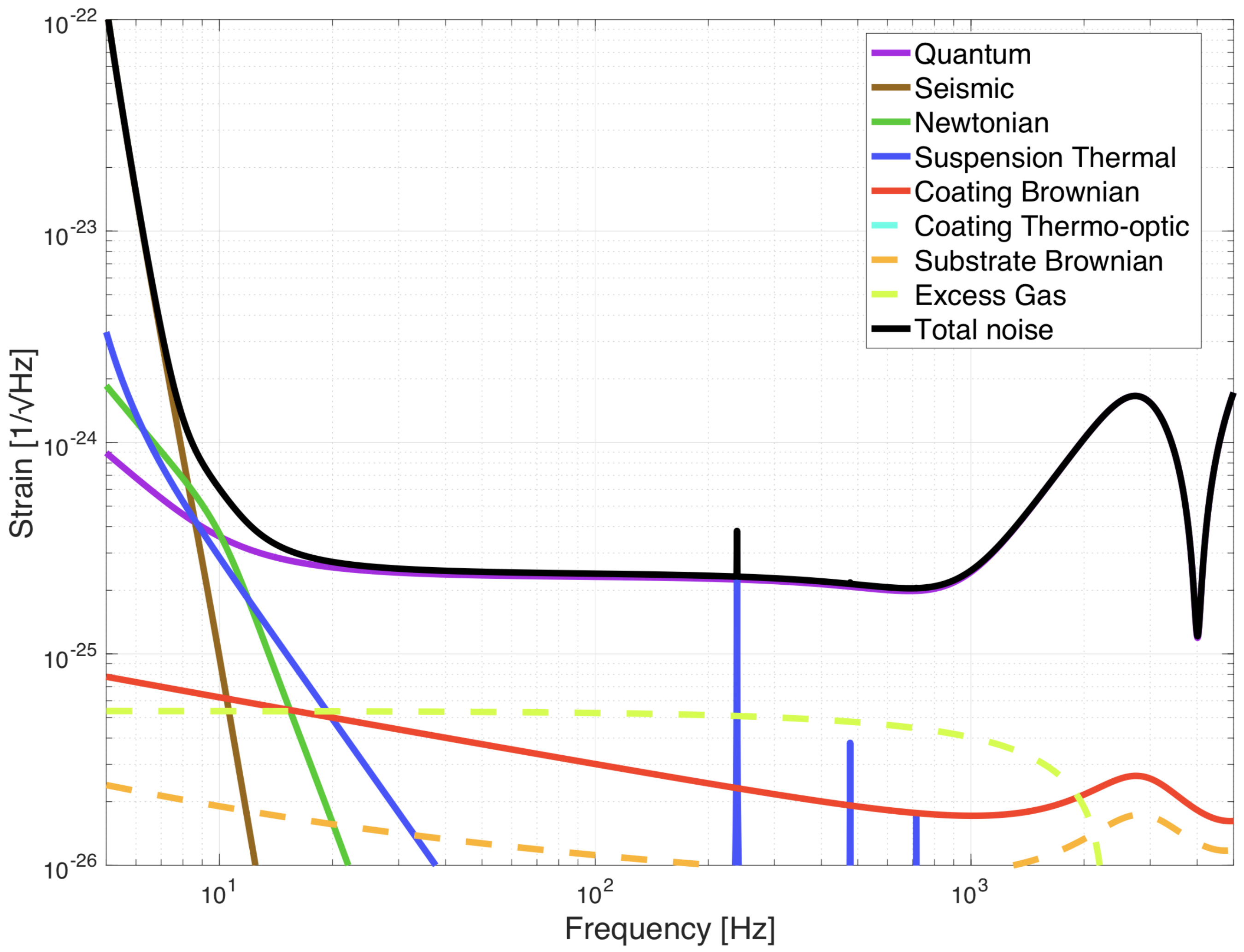}
	\caption{The figure summarizes the noise budget of the supernovae-optimized detector for a gravitational-wave signal with a 45 degrees tilt with respect to the arm cavities \cite{schilling1997angular}. Over the broad range of frequencies of interest, 500~Hz to 1500~Hz, the sensitivity is limited by quantum noise. The dip in sensitivity at 4~kHz corresponds to the pole of the signal recycling cavity.}
	\label{fig:SNOpt_noise}
\end{figure}

The optimal supernovae zero-detuned detector's noise budget is shown in Fig. \ref{fig:SNOpt_noise}. We find a longer signal recycling with a length of 180 m compared to 55m for Cosmic Explorer along with a transmissivity of the signal recycling cavity changed to 0.015 improved the detector's sensitivity by improving the quantum noise floor at higher frequencies. The loss in sensitivity around 3~kHz is due to the FSR of the arm cavity. The dip at 4~kHz corresponds to the pole of the signal recycling cavity. 

We also consider the effects of detuning the signal recycling cavity. We find detuning the signal recycling cavity with active compensation with the squeezing phase can be used to actively tune the third generation detectors in narrow bins of frequency without losing 15~dB of squeezing. It has been proposed that detuning the ground-based detectors can be useful in testing the general theory of relativity \cite{tso2018optimizing} with a joint operation with LISA \cite{danzmann1996lisa}. We will consider the applicability of these configurations to see if they provide any improvements for CCSN in section~\S\ref{sec:DeTune}.

\begin{figure}
    \centering
    \includegraphics[width=0.49\textwidth]{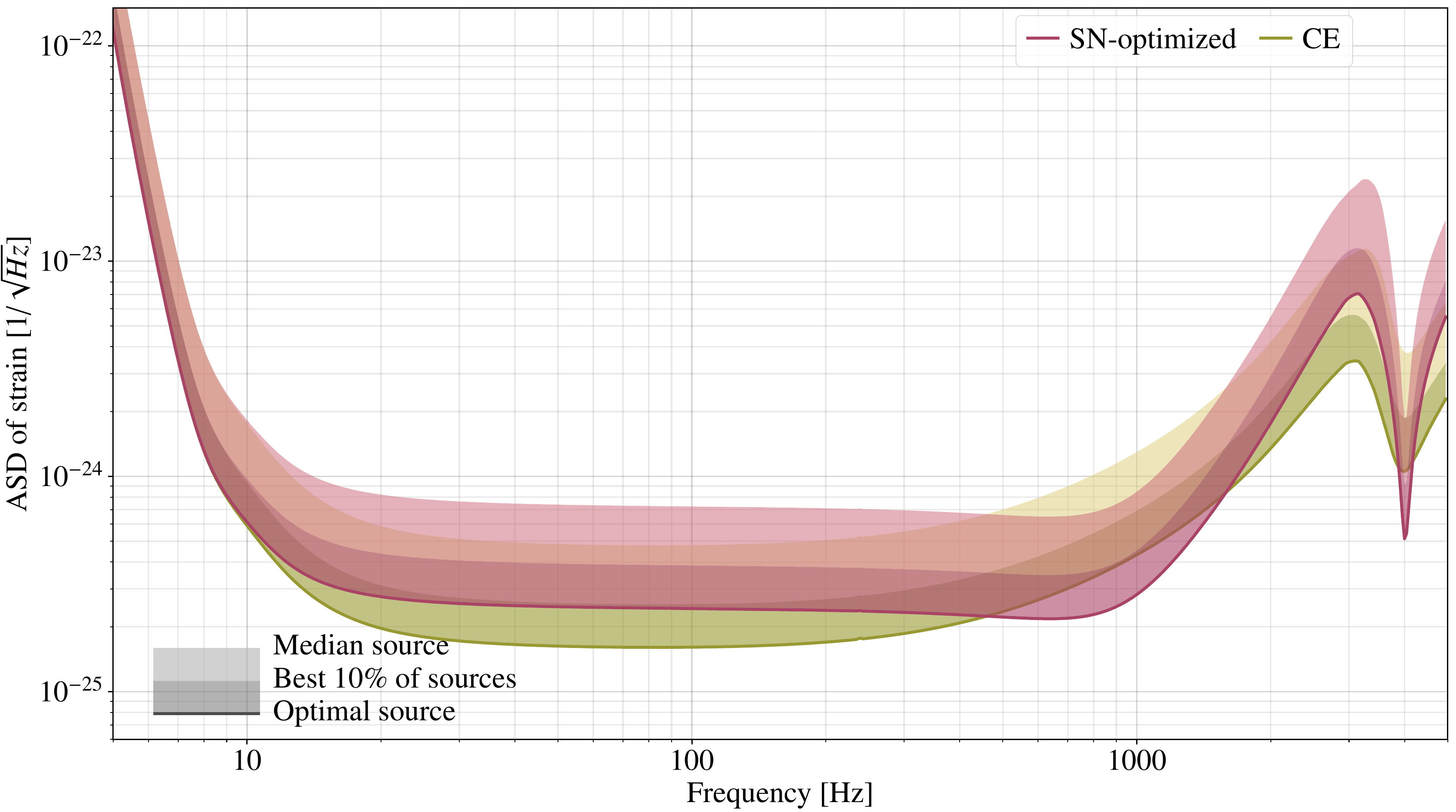}
    \caption{The figure summarizes the sky-averaged and orientation-averaged power spectral density of Cosmic Explorer and supernovae-tuned detector \cite{essick2017frequency}. We see that the Cosmic Explorer has a better noise floor from 10~Hz to 450~Hz. The supernovae-tuned detector has improved sensitivity over the range from 450~Hz to 1600~Hz. The numerical waveforms of CCSN suggest that a significant amount of power is emitted in this range. The optimization for CCSN improves the range from 70~kpc to 95~kpc for CCSN. However, this range improvement does not add any new galaxies. Therefore, the event rate does not change with the improved sensitivity and we are limited to sources within our galaxy.}
    \label{fig:BBSNDet}
\end{figure}

The optimization over the length of the signal recycling cavity and the transitivity of the signal recycling mirror to maximize the supernovae range with the phenomenological waveform in Fig. \ref{fig:ModelCCSN} leads to an improvement of approximately 30$\%$ in the range of CCSN as compared to the Cosmic Explorer design. However, extending the range from a 70 kpc to 95 kpc does not add any galaxies in our local universe. The optimized supernovae detector does not increase the detection rate as compared to the Cosmic Explorer. For the sources at a fixed distance, this corresponds to approximately 25$\%$ improvement in SNR.

The Fig. \ref{fig:BBSNDet} compares the broadband configuration of a zero detuned 40 km detector optimized for CCSN signals with the design of the Cosmic Explorer, both configurations have a 15dB squeezing. We improve on the sensitivity in the frequency range from 450~Hz to 1550~Hz at the cost of a loss in sensitivity from 10~Hz to 450~Hz. This results in a 15$\%$ loss in range for BNS. However, it still provides higher sensitivity for the post-merger signals based on the predicted frequencies of interest for post-merger oscillations \cite{abbott2017gw170817,bauswein2019identifying,bose2018neutron,takami2014constraining}. The table ~\ref{tab:DetSumry} summarizes the parameters and their corresponding ranges towards different gravitational-wave sources. One advantage offered by the supernovae optimized configuration is robustness. Without any squeezing, the supernovae optimized detector has a range extending to the LMC, whereas the range of the phenomenological SN waveform with Cosmic Explorer without squeezing is 32~kpc.

\begin{figure}
	\centering
	\includegraphics[width=0.49\textwidth]{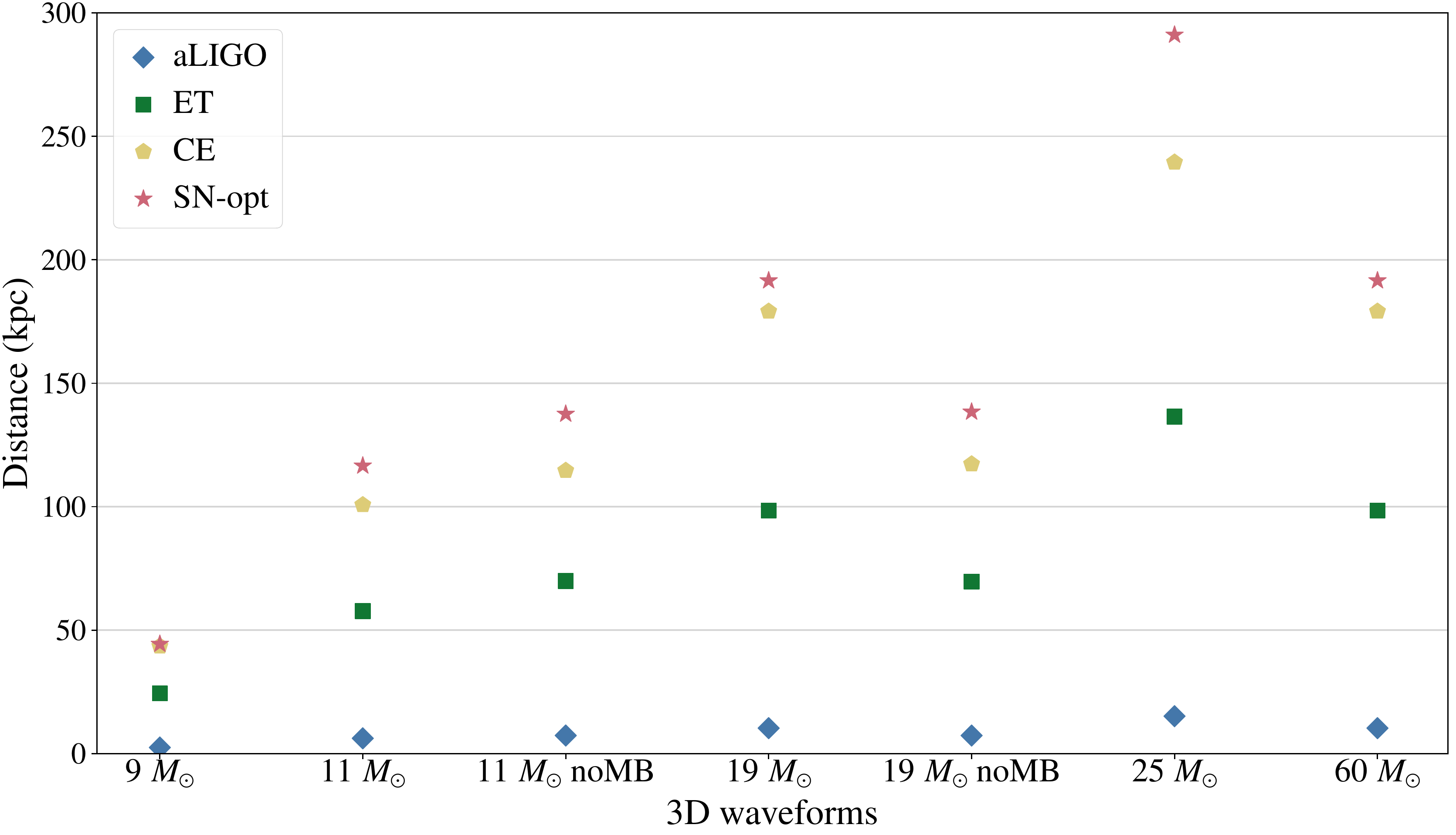}
	\caption{The figure summarizes the distance of the 3D waveforms for different second and third-generation gravitational wave detectors. We see for second-generation advanced LIGO detector that the optimal distances for the 3D numerical waveforms are limited to ~10kpc. The optimal distance is so small enough that we are not sensitive to all the galactic supernovae. All the third-generation detectors have optimal distance such that each detector is sensitive enough to detect gravitational waves from galactic CCSN. However, as evident from the plot above, for a source at a fixed distance, the ET will have the lower SNR as compared to Cosmic Explorer. The supernovae-optimized detector provides approximately a 25$\%$ improvement in the SNR as compared to Cosmic Explorer.}
	\label{fig:BB_WvfmSumry}
\end{figure}

Next, we use the noise curves of aLIGO, Cosmic Explorer, Einstein Telescope and Supernovae optimized detector configurations to compute the ranges for the 3D waveforms As stated earlier, the 3D waveforms are representative of astrophysically abundant stars which are not rapidly rotating and the corresponding gravitational wave strain emitted is small. Figure~\ref{fig:BB_WvfmSumry} summarizes the ranges of different waveforms based on their ZAMS mass. We see that the sensitivity of the third-generation of gravitational-wave detectors to CCSN is limited to sources within our galaxy. From the event rates of CCSN summarized in table \ref{tab:SNrate}, we find the corresponding event rate of observation of gravitational waves from CCSN (assuming a 100 $\%$ detector duty-cycle) is approximately one in fifty years.


\subsection{Detuning a large signal recycling cavity for narrow-band configurations}\label{sec:DeTune}
\begin{figure*}
    \centering
    \includegraphics[width=\textwidth]{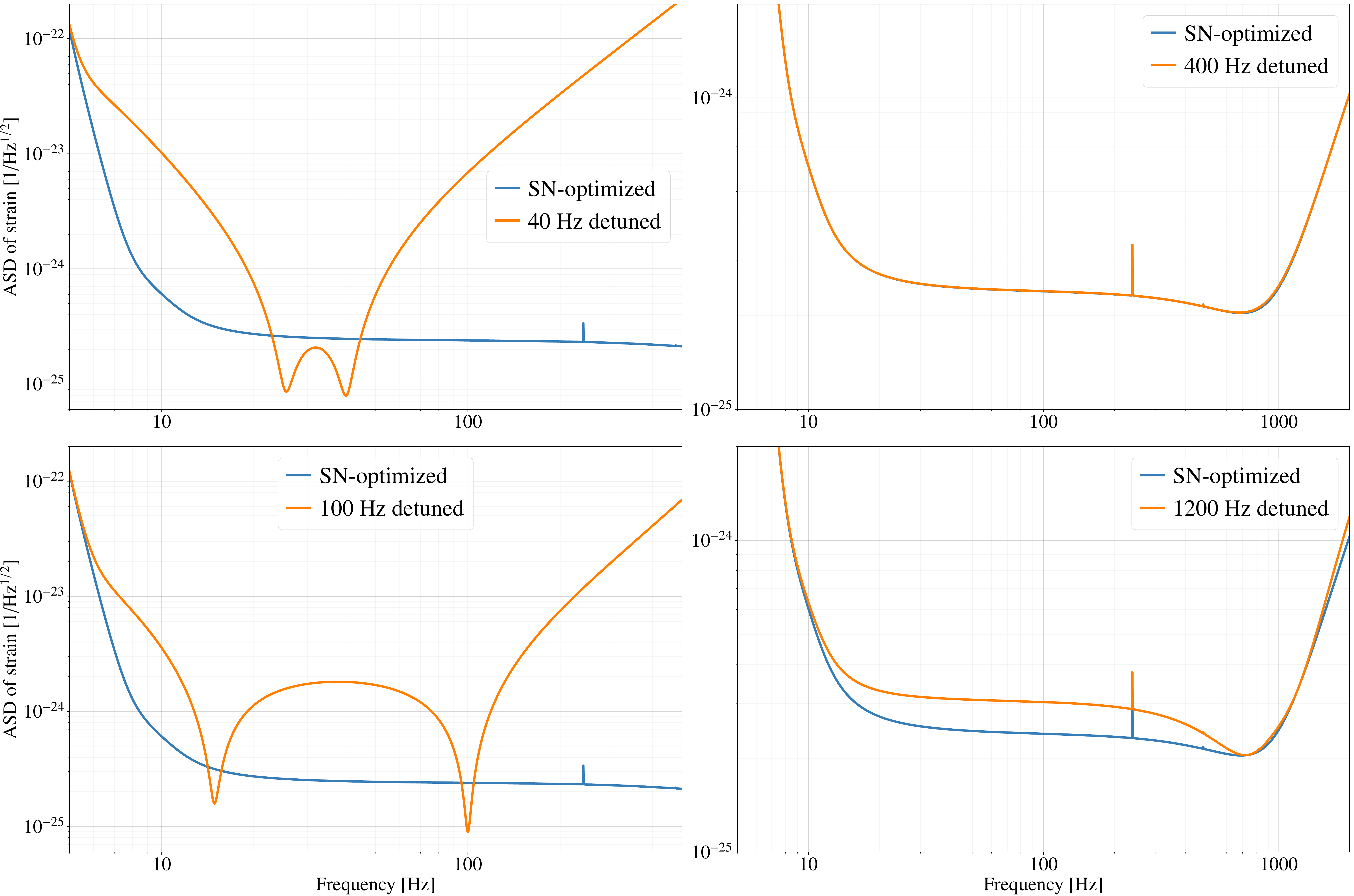}
    \caption{We explore the possibility of detuning the signal recycling cavity to improve the sensitivity towards CCSN. We find that detuning can be used to improve sensitivity in narrow bins of frequency below 400~Hz. This could, therefore, be used to study the ring-down modes of binary black-holes systems in collaboration with eLISA \cite{tso2018optimizing}. However, for improvements to the range of CCSN, this technique isn't useful.}
    \label{fig:DetunedRes}
\end{figure*}

A significant GW signal from CCSN lies in the frequency band from 500~Hz to 1500~Hz. The power emitted at different frequencies may vary depending on the astrophysical features of the star - mass, rotation speed, equation of state, etc \cite{Burrows2007_RapidRot, burrows2019three, Radice_2019_Characterstics, Heger_2005}. 

In this section, we do not change the detector parameters' such as the transitivity or the length of the signal recycling cavity. This is because these parameters cannot be changed once the detector design is laid out. However, one can detune the signal recycling cavity to maximize sensitivity in a narrow band of frequencies \cite{hild2007demonstration, ward2010length}. This response from detuning the signal recycling cavity arises from the two sidebands resonances in quantum noise \cite{BunnChen2004, buonanno2003quantum}. We consider the detuning of the signal recycling cavity at different frequencies. 

We maintain the frequency dependent squeezing of 15~dB. We achieve 15~dB squeezing in a detuned signal recycling cavity without losing the injected squeezing by actively changing the squeezing angle in accordance with the amount of detuning. Thus, detuning the signal recycling cavity along with actively changing the squeezing angle can be used to switch from a broadband zero-detuned detector to a narrow band detector with greater sensitivity for some frequencies determined by the magnitude of detuning. We perform another tier of optimization in which we actively vary the amount of detuning and the squeezing angle. We limit the amount of detuning in the range from $-\pi/5$ to $\pi/5$ and the squeezing phase is tuned in between $-\pi$ to $\pi$. To optimize the detector response at frequencies of 40~Hz, 100~Hz, 400~Hz and 1200~Hz, we inject a sine-Gaussian at each frequency and then maximize the range for this injected signal by varying only the detuning and squeezing angle for the supernovae optimized detector. 

We find that detuning the signal can improve the sensitivity of the detector in narrow bins of frequency below 400~Hz. We do not achieve improvements in sensitivity at higher frequencies therefore, we do not improve the range for different models by detuning the detector. There are no improvements in the optimal SNR values for a source at a fixed distance. In summary, detuning the signal recycling cavity is not useful for improving the Cosmic-Explorer-like detector's sensitivity to CCSN. Instead, detuning the signal recycling cavity at higher frequency degrades the sensitivity of the broadband supernovae-optimized detector. The corresponding results of detuning the signal recycling cavity are summarized in Fig. \ref{fig:DetunedRes}.

\subsection{Narrow-band Configurations tuned for Supernovae}\label{subsec:NBDetDes}
The parameters of the broadband supernovae-optimized detector were computed in section \S\ref{subsec:BB_CCSNDet} with two constraints. We will in this section relax those constraints and consider narrow-band detector configurations to maximize the range for CCSN. The phenomenological waveform we developed cannot be used for narrow-band optimization as the fit was performed to match the power of the 3D waveforms over a broad frequency bandwidth. Therefore, we find narrow-band configurations using a different technique.

The length of the signal recycling cavity can be changed to tune the resonant frequency of arising from the coupling of the signal recycling  cavity with the arms of the interferometer \cite{meers1988recycling, BunnChen2004}. The bandwidth of the resonance at a frequency $\omega_{r}$ is given by
\begin{equation}\label{eq:bw}
    B = \frac{\mathrm{cT}_{srm}}{4\mathrm{L}_{src}} 
\end{equation}
where $\mathrm{T}_{srm}$ is the transmissivity of the signal recycling mirror and $\mathrm{L}_{src}$ is the length of the signal recycling cavity. We choose the length of the signal recycling cavity at 150m, 300m and 750m are such that the resonant frequency $\omega_{r}$ is at 1000~Hz, 750~Hz and 500~Hz respectively. The equation \ref{eq:bw} is then inverted for bandwidth ranging from 250~Hz to 1600~Hz and the corresponding values of the transmissivity of the signal recycling mirror are calculated. 

\begin{figure}
	\centering
	\includegraphics[width=0.49\textwidth]{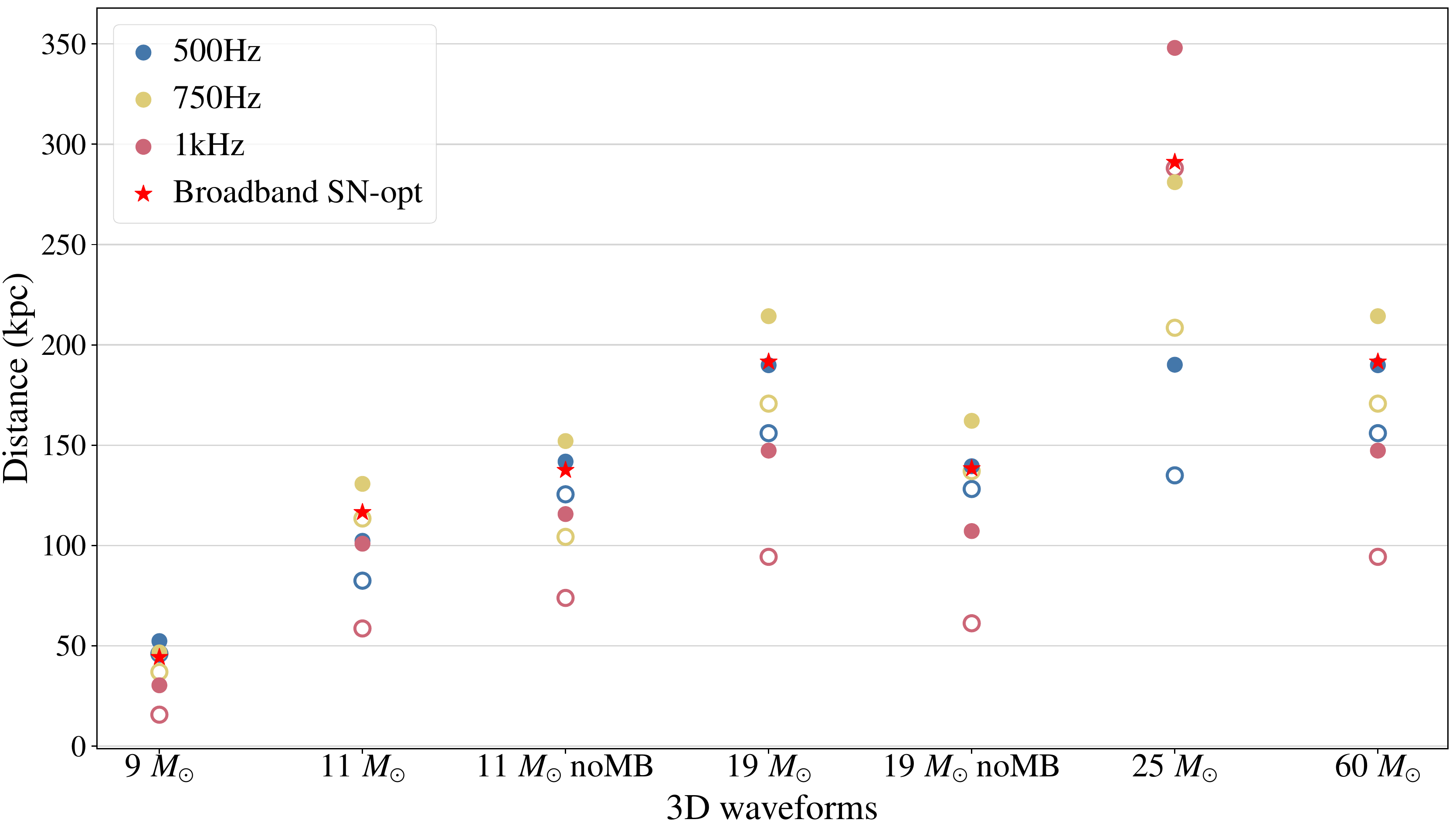}
	\caption{The figure summarizes the optimal distance of the different 3D waveforms for narrow-band detectors at frequencies 500~Hz, 750~Hz and 1000~Hz. The hollow circles denote the narrow-band detectors with a bandwidth of 250~Hz while the filled circles denote the bandwidth of 1600~Hz. The optimal distances from the broadband supernovae-optimized detector are represented as stars. We see tighter narrow-banding with a bandwidth of 250~Hz degrades the performance of the detector. The wider bandwidth of 1600~Hz around the 750~Hz narrow-band detector improves the optimal distances for most of the numerical waveforms.}
	\label{fig:NBdets}
\end{figure}

We find that a narrow-bandwidth of 250~Hz significantly affects the sensitivity of the detector towards CCSN. This is expected as we have stated earlier that the frequency spectrum of gravitational wave emission from CCSN is broadband. The range improvements achieved by narrow-band detectors at 500~Hz, 750~Hz and 1000~Hz with a bandwidth of 250~Hz are also varying from waveform to waveform and therefore is not model independent \ref{fig:NBdets}. When the bandwidth is increased to 1600~Hz the range improves for the 750~Hz narrow-band detector for some of the waveforms as shown in Fig. \ref{fig:NBdets}. The $\mathrm{L}_{src}$  = 300m and $\mathrm{T}_{srm}$ = 0.0064 give this narrow-band detector configuration. The mean improvement in optimal SNR with the 750~Hz narrow-band and 1600~Hz bandwidth detector is approximately 10~$\%$ with respect to the supernovae optimized broadband detector. However, we caution that the improvement from narrow banding is not the same across all the 3D numerical waveforms. Moreover, this comes at the cost of significant loss of sensitivity below 400~Hz and above 1100~Hz. The range for BNS drops to 3~Gpc (z=0.9) compared to 3.7~Gpc (z=1.1) for supernovae-optimized Cosmic Explorer and 4.3~Gpc (z=1.4) with respect to the Cosmic Explorer.

\section{Challenges in building a CCSN Detector to achieve Higher Event rates}\label{sec:FutureDet}
In section~\S\ref{subsec:BB_CCSNDet}, we find an optimized third-generation broadband gravitational wave detector for a CCSN signal has the range only to a few hundred kilo-parsec for the 3D numerical waveforms of CCSN.

\begin{figure}
	\centering
	\includegraphics[width=0.49\textwidth]{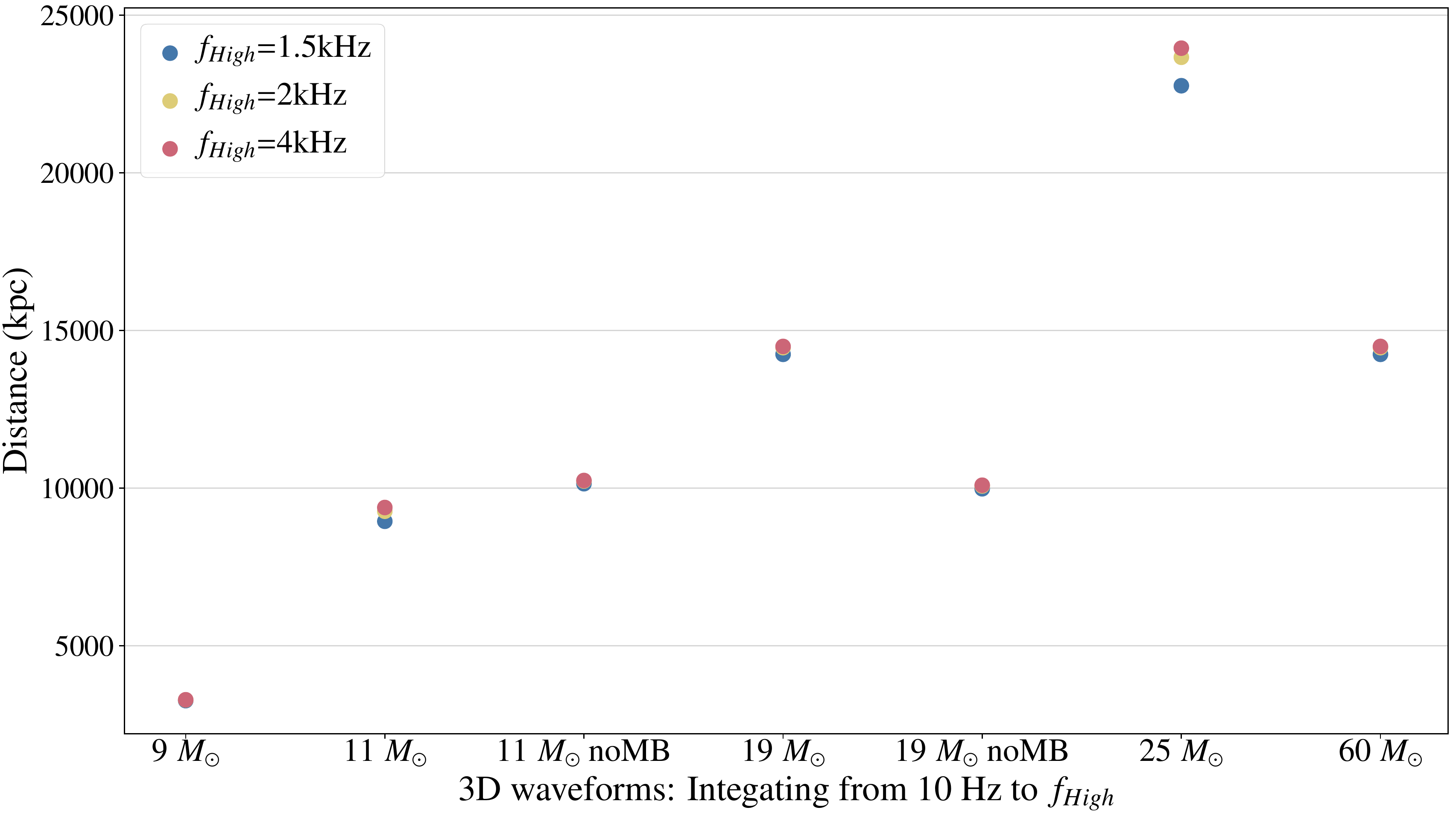}
	\includegraphics[width=0.49\textwidth]{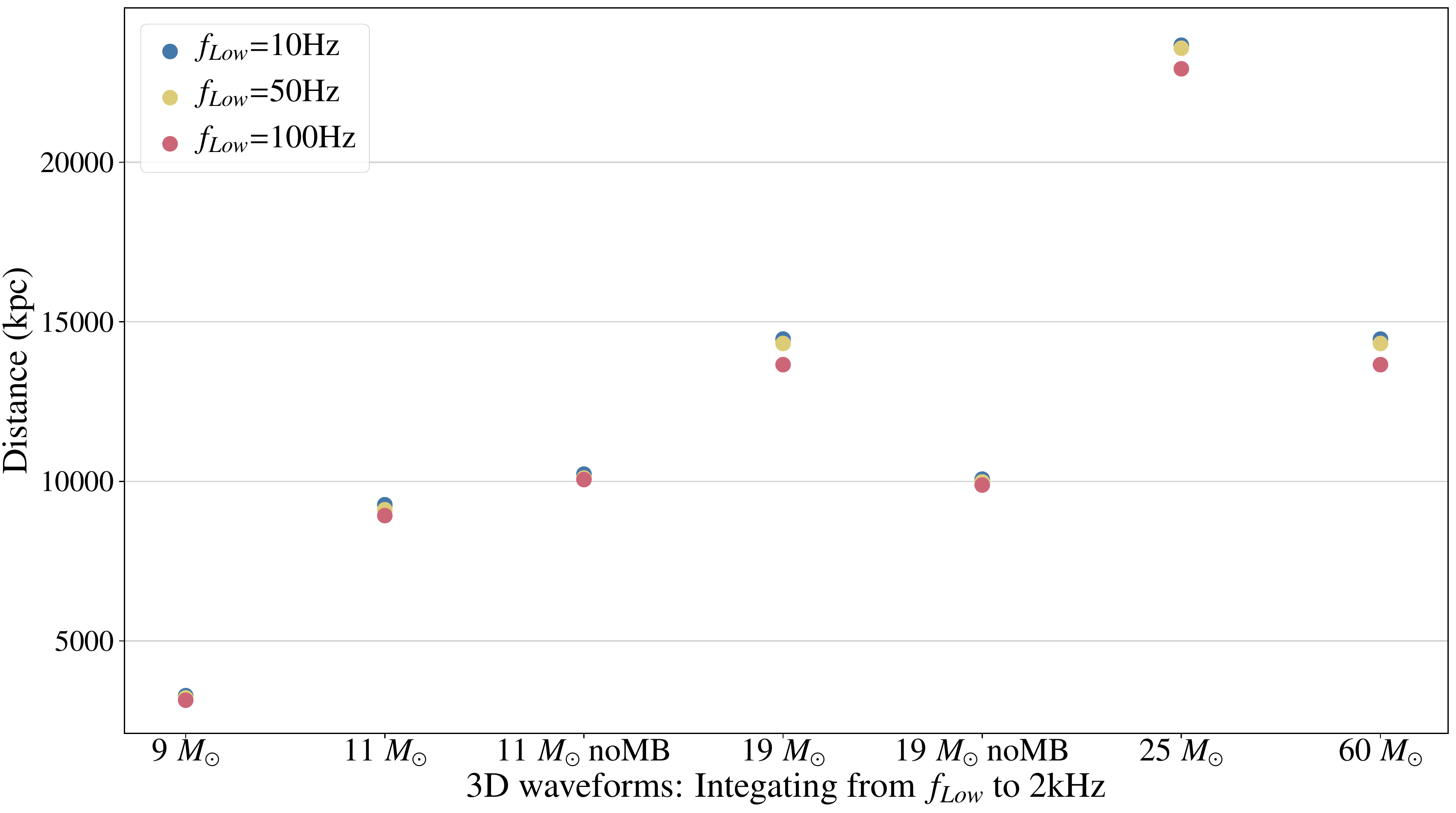}
	\caption{Considering toy detector with a flat PSD of 3$\times 10^{-27}~Hz^{-1/2}$ in range 10~Hz to $f_{High}$ (above) and $f_{low}$ to 2~kHz (below), the figure summarizes the range with the corresponding sensitivity and numerical waveform CCSN corresponding to their ZAMS mass. We see a broadband detector with a strain sensitivity of 3$\times 10^{-27}~Hz^{-1/2}$ from 200~Hz to 1.5~kHz is desired to achieve the ranges that would correspond to an observed event rate of one per year for gravitational-waves from CCSN.}
	\label{fig:StrainReq3D}
\end{figure}

We now address the question of what are the strain requirements for a gravitational-wave detector to be able to detect CCSN with an event rate of 0.5 per year. From the table \ref{tab:SNrate}, we see that this ``Hypothetical CCSN detector" must have a range of $\mathcal{O}$(10 Mpc) for CCSN to achieve an event rate of 0.5 per year. Moreover, a single detector, we need a signal to noise ratio (SNR) of 8 to define the detection of a signal against the background. Using the two constraints above we can calculate the minimum strain sensitivity required to achieve an event rate of 0.5 per year for the waveforms from 3D numerical simulations. The optimal distance for the numerical waveforms can be calculated by equation \ref{eq:optD}. The limits over the integral are defined by $f_{low}$ and $f_{high}$. To find the strain requirements for the different waveforms we assume a flat PSD over a broadband range of frequency ranging from $f_{low}$ and $f_{high}$. 
We consider two scenarios which are summarized in the figures \ref{fig:StrainReq3D}. First, we vary the upper limit of the frequency integrated -- $f_{High}$ with the lower limit of integration is held constant at 10~Hz. The second scenario where the upper limit of integration is constant at 2~kHz and we vary the lower frequency limit $f_{low}$. We find the minimum strain sensitivity required for the gravitational-wave detector to detect the CCSN with an event rate of 0.5 per year is $3\times10^{-27}~\mathrm{Hz}^{-1/2}$ over a frequency range of 100~Hz to 1500~Hz. 

Thus, we need a detector with sensitivity approximately a hundred times better than the Cosmic Explorer design to detect CCSN with an event rate of 0.5 per year. In the next section \S\ref{sec:FutureDet}, we will summarize the noise limitations of the third generation detectors and consider design parameters for gravitational-wave detectors beyond the scope of the third-generation to determine the technological hurdles to overcome in order to ever observe gravitational signals from CCSN more frequently.


\begin{figure}
	\centering
	\includegraphics[width=0.45\textwidth]{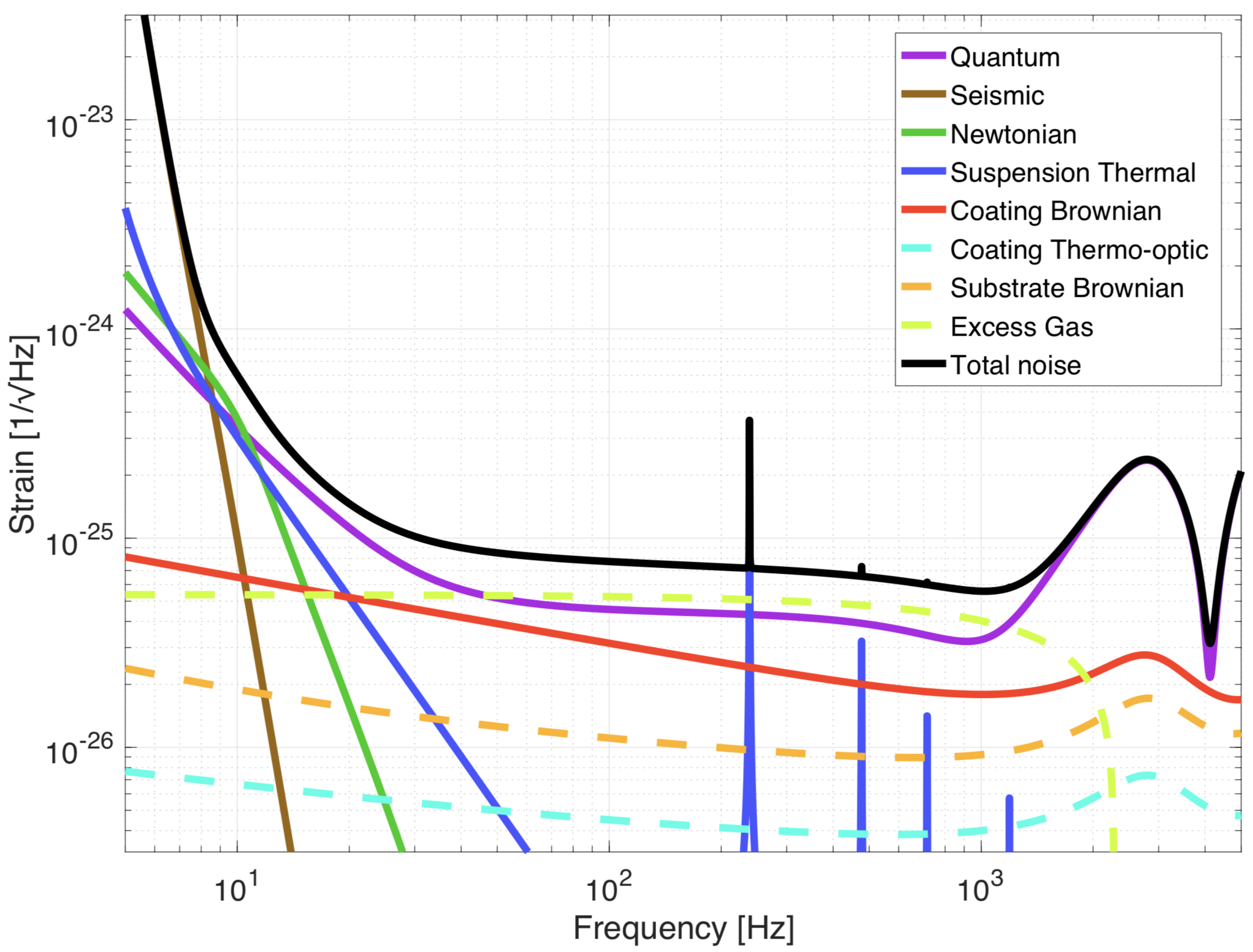}
	\includegraphics[width=0.45\textwidth]{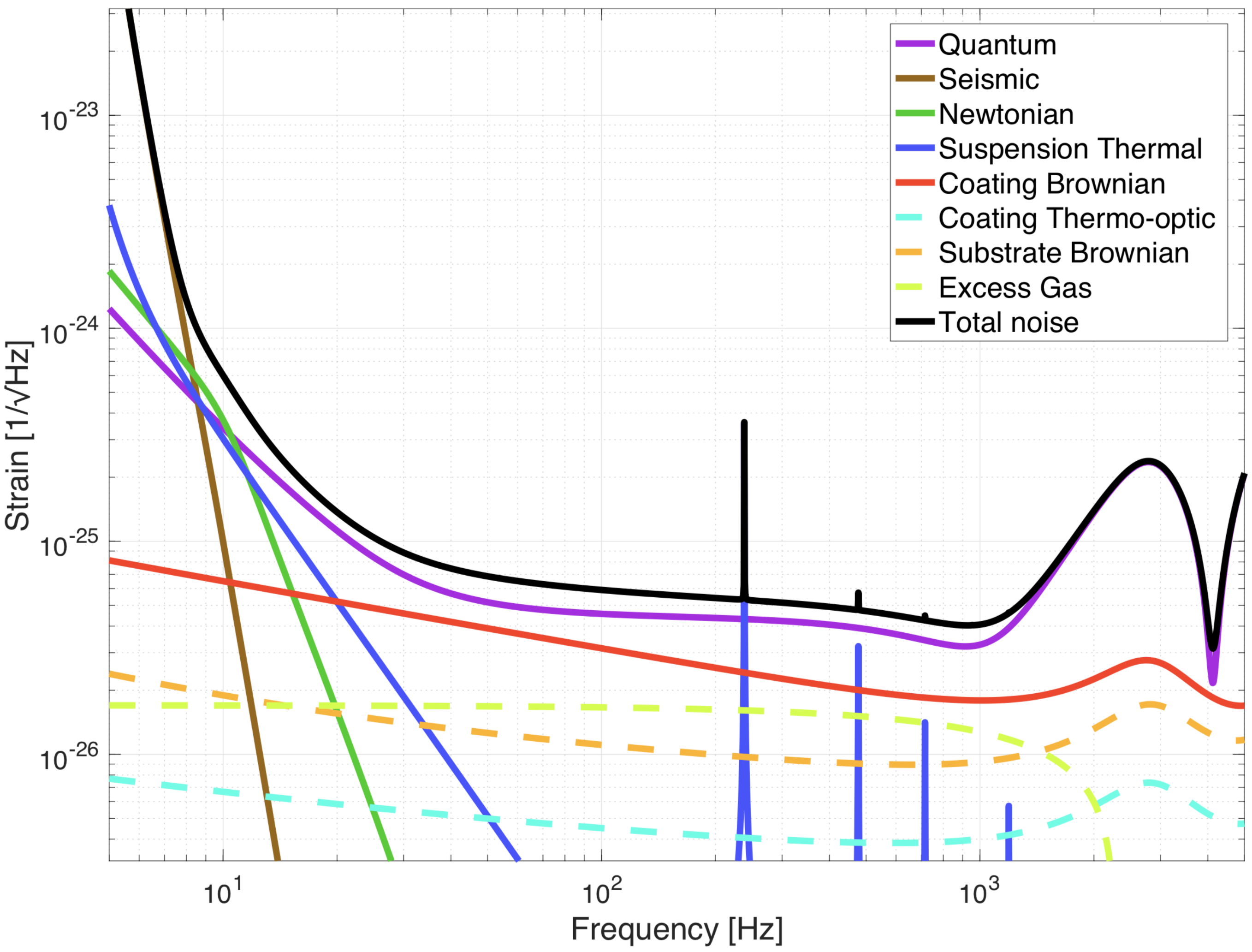}
	\includegraphics[width=0.45\textwidth]{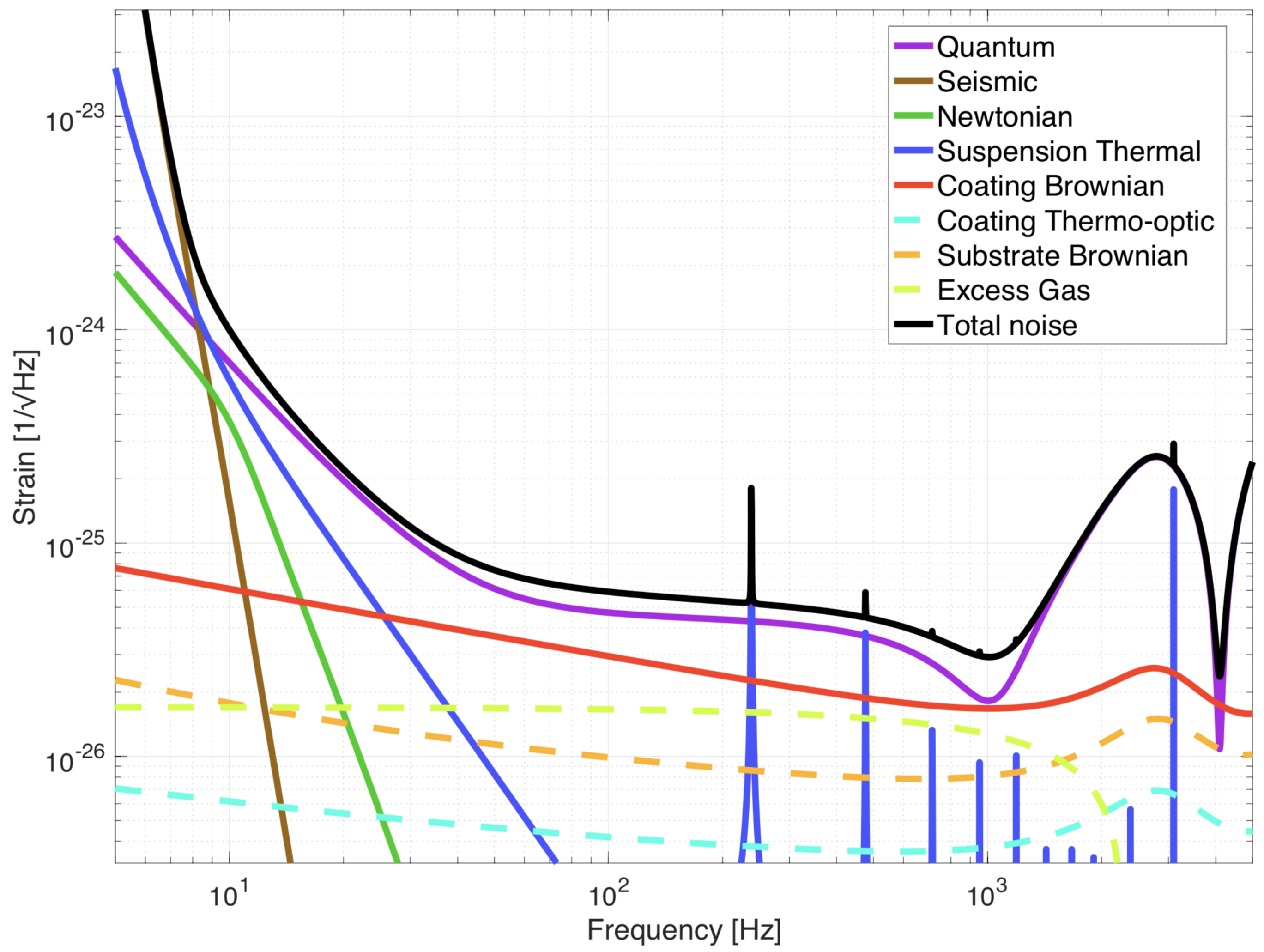}
	\caption{The figure above summarizes the noise budgets for the Hypothetical detector configurations. We see from the figure on the top that the detector's sensitivity is limited by residual gas noise. Therefore, we reduce the residual gas pressure by a factor of ten from CE design. The plot in the middle and bottom plots show optimization results without changing the transmittance of the power recycling cavity and with active changes in the transmittance of the power recycling cavity. Thereby, changing the gain of the power recycling cavity and the finesse of the detector. We will refer to the two detector configurations as Hypothetical-1 and Hypothetical-2 respectively.}
	\label{fig:3GplusNoises}
\end{figure}


It is evident from Fig. \ref{fig:SNOpt_noise} that the sensitivity is limited by the quantum noise in the broad range of frequencies. The standard quantum noise limit is dependent primarily on the length of the arm cavities, the test masses and the power of the input laser \cite{braginsky1996quantum}. The length of the arm cavities cannot be increased any further as the $\mathrm{f}_{FSR}$ would significantly affect the performance of the detector at the frequencies of interest. As a result, we set the length of the Hypothetical detectors to 40~kms. Increasing the power of the input laser is the one possibility to reduce quantum noise. We assume an input laser power of 500W. At high frequencies, the quantum noise in the detector manifests itself as shot noise and is limited by photon number arriving at the photo-detector. To see the best we can achieve, we set the photo-detection efficiency of the photo-detector in Hypothetical to 1 (from 0.96 for CE design). For the same reason, we also set the optical and squeezing injection losses in the detector to zero. 

\begin{figure}
	\centering
	\includegraphics[width=0.49\textwidth]{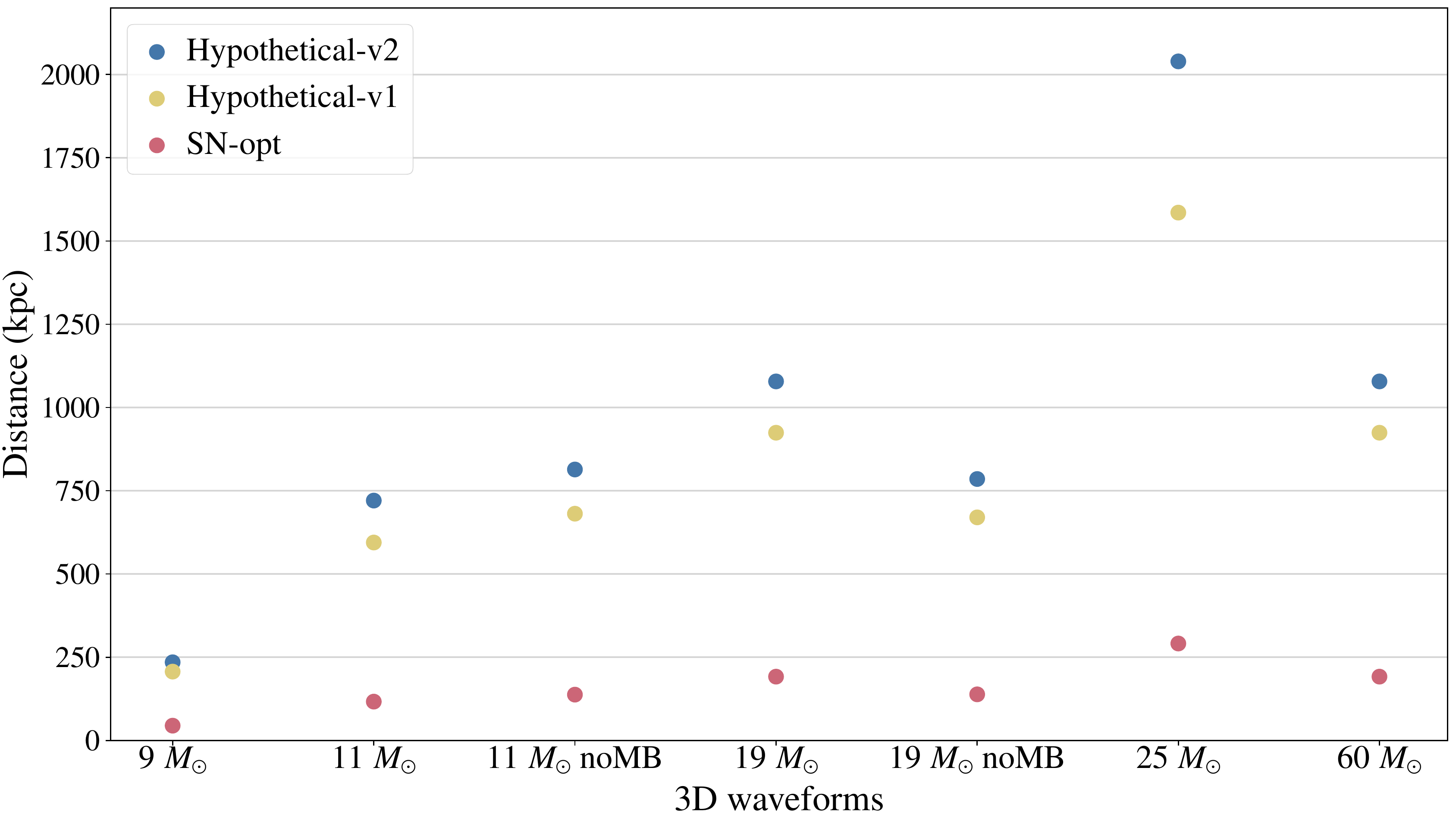}
	\caption{The plot shows with extreme technological upgrades to the third-generation detectors discussed in section \S\ref{sec:FutureDet}, we optimal distances for the CCSN is limited to 1Mpc. The event rate for the observation of gravitational waves from CCSN is still low but improves to one in twenty years. }
	\label{fig:3GplusRanges}
\end{figure}

The coating thermal noise and the residual gas noise are the next limiting factor in the system. We reduce the substrate absorption by an order of magnitude from CE design. Lastly, as the frequency range of interest is from 100Hz we can sacrifice the sensitivity at lower frequencies. Thus, we can reduce the masses of the mirrors as we are interested in improving the shot noise characteristics of the detector, at the cost of higher radiation pressure noise. In this setup we optimize over the length of the signal recycling cavity $\mathrm{L}_{src}$, the transmissivity of the signal recycling mirror $\mathrm{T}_{srm}$, the transmissivity of the input test mass $\mathrm{T}_{itm}$ and the scale mass parameter to change the masses of the mirror. The optimization over these parameters is aimed at maximizing the range for the representative supernovae waveform, we will reference this optimized detector as {\it Hypothetical-1}. 

The quantum noise limit in a dual-recycled Fabry-Perot interferometer also depends on the gain of the power recycling cavity \cite{BunnChen2004,buonanno2003quantum}. We will in another independent optimization also tune the transmissivity of the power recycling mirror $\mathrm{T}_{prm}$ along with the above parameters. We define this supernovae-optimized detector as {\it Hypothetical-2}. The table \ref{tab:DetSumry} summarizes the optimal parameters of different detectors. Fig. \ref{fig:3GplusNoises} shows the noise budget of the Supernovae optimized Hypothetical detectors. We see that the residual is the limiting source of the noise. Removing the residual gas noise improves the noise floor of the detector by a factor of two in the wide range of frequencies of interest, see Fig. \ref{fig:3GplusNoises}. After removing the residual gas noise, we are limited in sensitivity by quantum noise over the broad range of frequencies.

The strain sensitivity achieved after removing the residual gas noise is $5\times10^{-26} \mathrm{Hz}^{-1/2}$. The improvements in photo-detection efficiency, the input laser power, substrate coatings and minimization of optical losses are not sufficient to achieve a strain sensitivity of the order of $3\times10^{-27}~\mathrm{Hz}^{-1/2}$ required to detect CCSN with an event rate of one in two years (see section \S\ref{sec:FutureDet}). 

Lastly, we revisit the numerical waveforms of core-collapse supernovae to see the ranges achieved by the Hypothetical supernovae-optimized detector designs. We find for the 3D waveforms from numerical simulations have a mean distance of 800~kpc, see Fig. \ref{fig:3GplusRanges}. Thus, with beyond the third generation detector designs, we would be able to observe core-collapse supernovae from Andromeda. The corresponding event rate is of the order of one in twenty years. The event rate calculation assumes a 100$\%$ duty cycle of the detector. The observation rate of gravitational waves from CCSN is low even for gravitational-wave detectors beyond the scope of the third-generation detectors.


\begin{table*}
  	\begin{center}
  	\begin{tabular}{|p{4cm}|p{2.5cm}|p{2.7cm}|p{2.5cm}|p{2.5cm}|p{2.5cm}|}
  		\hline
        \ Parameters & aLIGO & Cosmic Explorer-2 & SN Optimized & Hypothetical-1 & Hypothetical-2 \\
        \hline
        \ Input Power       & 125W  & 220W & 220W & 500W & 500W  \\
        \ SRM transmission  & 0.325 & 0.04  & 0.015 & 0.0030 & 0.0122\\
        \ ITM transmission  & 0.014 & 0.014 & 0.014 & 0.0036 & 0.0269\\
        \ PRM transmission  & 0.030 & 0.030 & 0.030 & 0.030  & 0.0011\\
        \ $\mathrm{L}_{src}$         & 55m & 55m & 175m & 30m & 260m \\
        \ Finesse           & 446.25 & 447.52 & 447.52 & 1745.33 & 233.33 \\
        \ Power Recycling Factor & 40.66    & 65.32 & 65.32 & 94.25 & 1300.09 \\
        \ Arm power         & 712.43 kW     & 2025.70 kW & 2025.70 kW & 26.06 MW & 47.61 MW \\
        \ Thermal load on ITM & 0.386 W     & 1.150 W & 1.150 W & 13.094 W & 24.180 W\\
        \ Thermal load on BS & 0.051 W      & 0.253 W & 0.253 W & 0.008 W & 0.080 W\\
        \ BNS range         & 173.00 Mpc    & 4.29 Gpc & 3.67 Gpc & 5.32 Gpc & 5.09 Gpc \\
        \ BNS horizon       & 394.83 Mpc    & 11.05 Gpc & 9.49 Gpc & 12.97 Gpc & 12.53 Gpc \\
        \ BNS reach         & 246.06 Mpc    & 8.54 Gpc & 6.90 Gpc & 11.56 Gpc & 10.80 Gpc \\
        \ BBH range         & 1.61 Gpc      & 6.13 Gpc & 6.10 Gpc & 6.15 Gpc & 6.09 Gpc \\
        \ BBH horizon       & 3.81 Gpc      & 11.86 Gpc & 11.85 Gpc & 11.85 Gpc & 11.70 Gpc \\
        \ BBH reach         & 2.54 Gpc      & 11.73 Gpc & 11.73 Gpc & 11.72 Gpc & 11.52 Gpc \\
        \ Supernovae range     & 4.34 kpc   & 71.95 kpc & 94.24 kpc & 540.53 kpc & 716.03 kpc \\
        \ Supernovae horizon   & 9.84 kpc   & 163.08 kpc & 213.61 kpc & 1225.22 kpc & 1623.06 kpc \\
        \ Supernovae reach     & 6.10 kpc   & 101.04 kpc & 132.35 kpc & 759.15 kpc & 1005.65 kpc \\
        \ Stochastic Omega     & 2.36e-09   & 1.82e-13 & 2.77e-13 & 1.1e-13 &  2.58e-13 \\
        \hline
  	    \end{tabular}
      \end{center}
    \caption{ Summary of All Detectors}
    \label{tab:DetSumry}
\end{table*}

\section{Conclusion}\label{sec:con}
We have shown that it is possible to tune a Cosmic Explorer detector to increase the range to CCSN by approximately 25$\%$. This range improvement does not translate to increase in detection rate due to the inhomogeneity of the local universe. Therefore, even optimized third-generation gravitational-wave detectors will be limited to CCSN sources within our galaxy and the Magellanic Clouds. Assuming the detectors have a duty-cycle of 100$\%$ the corresponding event rate of CCSN is one in fifty years. Incorporating the detector downtime and duty-cycle would further decrease the event rate of observed gravitational-wave signals from CCSN.

However, if such an event were to occur, the broadband supernovae-optimized detector would improve the SNR by of sources by 25$\%$. This improvement would facilitate help understand the properties of the progenitor star in the rare event of CCSN observation. The supernovae-optimized detector has a slightly reduced sensitivity to the inspiral of neutron stars, but the high-frequency improvements would benefit the study of post-merger signatures and the late-time behavior of the inspiral. 


We find that a gravitational-wave detector would require a strain sensitivity of the order of 3$\times$10$^{-27}$ $\mathrm{Hz}^{-1/2}$, over a frequency range from 100~Hz to 1500~Hz in order to guarantee a high rate of CCSN detection. At this strain sensitivity, as per the current estimates of the BNS background, the stochastic background from BNS mergers would contribute as the fundamental sources of noise \cite{abbott2018gw170817}. This along with technological challenges discussed in section \S\ref{sec:FutureDet} poses significant hurdles in achieving an event rate of one per year for the observation of gravitational-waves from CCSN based on the present models and knowledge of gravitational-wave emission from CCSN. The technological requirements for these upgrades are beyond the requirements for the third-generation detector. With drastic improvements of an input laser power of 500~W and a photo-detection efficiency of 1, an order of magnitude improvement in the residual gas noise and coating noise from the Cosmic Explorer design, and assuming minimal optical losses in {\it Hypothetical} detectors. We find that after optimizing these detector configurations to maximize for the supernovae range the range extends to Andromeda for some of the CCSN numerical waveforms. The event rate achieved with such a hypothetical detector is one in twenty years. 


\section{Acknowledgements}
We would like to thank Evan Hall and Joshua Smith for their help with \texttt{GWINC}. We thank Geoffrey Lovelace and Christopher Wipf for helpful discussions in regards with the astrophysical implications of the results. VS, SB, DAB, and CA thank the National Science Foundation for support through award PHY-1836702. AB, DR, and DV acknowledge support from the U.S. Department of Energy Office of Science and the Office of Advanced Scientific Computing Research via the Scientific Discovery through Advanced Computing (SciDAC4) program and Grant DE-SC0018297 (subaward 00009650), the U.S. NSF under Grants AST-1714267 and PHY-1144374, the DOE/ASCR INCITE program under Contract DE-AC02-06CH11357, a Blue Waters PRAC (under OCI-0725070, OAC-1809073, and ACI-1238993), and the National Energy Research Scientific Computing Center (NERSC) under contract DE-AC03-76SF00098. DAB thanks NSF award PHY-1748958 to the Kavli Institute for Theoretical Physics for support.

\bibliographystyle{apsrev4-1}
\bibliography{main}

\end{document}